\documentclass[12pt]{article}

\usepackage{newtxtext,newtxmath}

\usepackage{graphicx}

\usepackage[letterpaper,margin=1in]{geometry}

\renewenvironment{abstract}
	{\quotation}
	{\endquotation}

\date{}

\makeatletter
\renewcommand{\fnum@figure}{\textbf{Figure \thefigure}}
\renewcommand{\fnum@table}{\textbf{Table \thetable}}
\makeatother

\usepackage{scicite}

\usepackage{booktabs} 

\usepackage{url}

\def\scititle{
	Ablation-Driven Selective Laser Direct Mass Transfer for Micro-LEDs with High Speed, Enhanced Gap Tolerance, and High Precision
}
\title{\bfseries \boldmath \scititle}

\author{
	Yuxuan~Cao$^{1,2}$,
	Yulong~Wang$^{1}$,
	Jürgen~Brugger$^{2,\ast}$,
	Zhen~Zhang$^{1,3,\ast}$\and
	\small$^{1}$Beijing Key Laboratory of Transformative High-end Manufacturing Equipment and Technology, \\
	\small Department of Mechanical Engineering, Tsinghua University, Beijing 100084, China.\and	
	\small$^{2}$Microsystems Laboratory, Ecole Polytechnique Federale de Lausanne
(EPFL), 1015 Lausanne, Switzerland.\and
	\small$^{3}$State Key Laboratory of Tribology in Advanced Equipment, Tsinghua University, Beijing 100084, China.\and
	\small$^\ast$Corresponding authors. Email: juergen.brugger@epfl.ch (J.B.), zzhang@tsinghua.edu.cn(Z.Z.)
}

\begin{document} 

\maketitle

\begin{abstract} \bfseries \boldmath

Micro light-emitting diodes (micro-LEDs) offer superior performance for next-generation displays, yet their scalable manufacturing is severely bottlenecked by the mass transfer process from donor to receiver substrates. The existing laser-induced forward transfer (LIFT) methods are restricted by low throughput, limited selectivity, and long reaction times. In this study, we introduce a novel defocused direct laser ablation transfer (D-DLAT) method utilizing polyimide as a dynamic release layer. 
{Whereas most indirect laser transfer methods trigger chip release followed by passive descent toward the receiver, D-DLAT directly ablates the release layer to actively propel the small-sized chips towards the receiver via ablation shock.}
By optimizing the ablation profiles and scanning paths, this method achieves high-speed transfer with significantly enhanced precision and  robustness. We demonstrate the successful transfer of micro-LEDs ranging from 600 µm down to 35 µm in size, 
{with a minimum reported placement error of less than $1$ \textmu m},
and a critical transfer gap of up to 50 µm. Furthermore, the D-DLAT approach enables full selectivity, facilitating complex patterned transfers and array repitching operations that are essential for full-color display integration. Our findings indicate that D-DLAT effectively overcomes the fundamental limitations of current transfer technologies, demonstrating substantial potential for the scalable and high-yield manufacturing of advanced micro-LED displays. Ultimately, D-DLAT establishes a robust paradigm for the high-speed manipulation and heterogeneous integration of micro-optoelectronic devices, extending its profound implications beyond displays to the broader landscape of advanced micro-assembly and flexible electronics.

\end{abstract}


\subsection*{INTRODUCTION}

Micro light-emitting diodes (micro-LEDs) have emerged as a promising technology for next-generation displays\cite{kim2025future_1dzlyb, Wu2020Full-Color}. Compared with conventional display technologies such as liquid crystal displays (LCDs) and organic light-emitting diodes (OLEDs), micro-LEDs offer significant advantages in performance. These advantages include high brightness ($>$100,000 nits), low energy consumption (10\% of that of LCDs and 50\% of that of OLEDs), rapid response times ($<$1 \textmu s), and durable lifespan ($>$100,000 h) owing to the inorganic nature of the devices\cite{gao2024microsecondr_1ka9c4, wu2024ultrahigh_qg39s9, Anwar2022Recent, ding2019micro}. These attributes make micro-LEDs highly suitable for high-performance display applications, such as mixed reality (MR) devices, wearable electronics, and high-dynamic-range (HDR) displays\cite{sim2025microdisplay_1k386e, park2026_1uo3ph}.

The manufacturing of micro-LED displays begins with the fabrication of monochromatic chips at a high density on their respective source wafers. Subsequently, chips of the three primary colors (red, green, and blue) are selectively transferred and re-pitched from their respective source wafers (also known as donors) onto a unified target substrate (also known as a receiver)\cite{lee2023fluidic_15doom, zhou2020growth_sniib7}. This transfer process, known as ``mass transfer'', requires the precise placement of tens of millions of micro-LED chips onto the target substrate with high accuracy\cite{luo2026laserassiste_157mkc}. It remains the most critical and challenging step in micro-LED display fabrication and has yet to be fully resolved\cite{hwang2023waferscale_1csq5f}.

The fundamental requirements for the mass transfer process are summarized as follows:

\begin{enumerate}
\item High precision: The overall displacement during transfer must not exceed 10\% of the dimension of the chip. Otherwise, the device faces the risk of failing to establish a proper electrical connection, which results in an unilluminated pixel on the display\cite{tenpercentthesis}.
\item High throughput: The transfer reaction time should be minimized to maximize the overall efficiency of production.
\item Full selectivity: The process must be capable of transferring any specific chips to any designated positions while leaving the remaining chips on the donor unaffected. This capability enables the integration of chips from different donors at the required pitch and pattern to construct full-color displays.
\end{enumerate}

To address the challenges of mass transfer, various techniques have been developed, with each presenting distinct advantages and limitations. 
The mechanical pick-and-place method utilizes robotic arms or micro-grippers to individually pick up and place micro-LED chips onto the target substrate. This approach offers high precision and selectivity, which allows for the transfer of specific chips to the desired locations. However, the throughput of this method is limited owing to the sequential nature of the process, which renders the technique less suitable for large-scale production\cite{zhang2009autonomous,peng2011analysis}.
The elastomer stamp transfer method employs a soft polymer stamp to retrieve and release micro-LED chips through adhesion forces\cite{cok2017inorganic, li2024manufacturing, kim2019control, lu2020optimization}. Although this method provides high precision, the technique is constrained by a relatively low throughput. Furthermore, as a contact transfer technique, this approach can only replicate the entire pattern of chips from the donor to the receiver, demonstrating no selectivity. Subsequent studies introduce improvements to transfer exclusively the selected chips through the use of patterned stamps; nevertheless, free repitching remains unachievable owing to the contact nature of the process and the stamp design\cite{zhao2022smart, zhao2023magnetically}.
The fluidic self-assembly method immerses the chips in a fluidic solution and guides the devices to fill the slots on the receiver substrate through surface tension and gravity. This method achieves high throughput but demonstrates limited precision. Additionally, to distinguish the chips of the three primary colors, complex designs for the shape of the chips and the slots are required, which further degrades the overall yield and reliability of the system\cite{chang2023concurrent, lee2023fluidic, cho2019fluidic, ji2022alternative}.

Laser-assisted transfer methods have garnered significant attention owing to the potential for high throughput and selectivity\cite{Serra2018Laser‐Induced,  rehman2023fundamentals}. High throughput is realized through high-speed laser scanning or the application of parallel laser pulses, whereas selectivity is achieved via the steering of the laser beam. Nevertheless, ensuring the precision of laser-assisted methods remains a substantial challenge, which necessitates the careful design of the laser interaction mechanism. Numerous prior investigations have demonstrated the transfer of liquids, metal films, and other materials through a process designated as the laser-induced forward transfer (LIFT) technique\cite{banks2008femtosecond, banks2008triazene, stewart2010laser, yang2024laser}.

The LIFT technique typically involves the incorporation of a sacrificial layer (commonly referred to as a dynamic release layer, or DRL) between the donor substrate and the components designated for transfer (commonly referred to as the flyer). Upon irradiation by a laser pulse, the DRL undergoes rapid decomposition or deformation, which enables the transfer of the flyer. This concept has been adapted for the transfer of micro-LEDs by various research groups\cite{Miller_Marinov_Swenson_Chen_Semler_2012, huang2025high, li2024laser, luo2024switchable, gai2026gradient, wang2020programmable, chen2023laser, xie2024high, yang2024super}; however, all the existing implementations exhibit specific limitations to date.

The existing LIFT methods for micro-LEDs can be categorized into two types based on the laser interation mechanism: the direct laser ablation method, which utilizes a laser to directly ablate and remove the DRL to release the chips from the donor; and the indirect laser transfer method, which utilizes a laser to induce deformation, alterations in adhesion, or other physical effects within the DRL to release the chips. The majority of prior studies fall into the latter category.
In 2013, the \textit{tm}SLADT method \cite{Miller_Marinov_Swenson_Chen_Semler_2012} first configured a multi-layer DRL that generates a blister upon laser irradiation to push the micro-LED chips from the donor. Silicon dies with dimensions of 600 \textmu m were transferred with a success rate of 65\% and a displacement of 61.2 \textmu m. Subsequent studies introduced improvements to the blister-actuated LIFT method, achieving a minimum chip size of 35 \textmu m \cite{huang2025high}, a minimum placement error of 2 \textmu m \cite{huang2025high}, a transfer gap of 25 \textmu m \cite{huang2025high}, and a minimum response time of 20 ms \cite{li2024laser}. Other investigations \cite{luo2024switchable, gai2026gradient} utilized infrared lasers to heat the polymer and induce deformation within the DRL, which subsequently releases the chips. These approaches demonstrated a minimum chip size of 100 \textmu m \cite{gai2026gradient}, a precision of 5 \textmu m \cite{gai2026gradient}, and a response time of 10 ms \cite{luo2024switchable}. In 2020, Wang et al. \cite{wang2020programmable} employed thermal release tape (TRT) as the DRL and utilized a laser to heat the TRT, which causes the material to lose adhesion and release the chips onto a polydimethylsiloxane (PDMS) receiver. However, as a contact transfer technique, this approach provides only limited selectivity. In 2023, Chen et al. \cite{chen2023laser} introduced further improvements by utilizing TRT in conjunction with polyimide (PI) as the DRL. They designed a two-step laser irradiation process to deform the DRL, reduce the adhesion, and thereby release the chips. This method achieved a minimum chip size of 150 \textmu m, a precision of 4 \textmu m, a transfer gap of 65 \textmu m, and a response time of 200 ms.

{In the indirect-transfer methods discussed above, the chips are released from the donor and passively descend toward the receiver. Among these methods, chip release requires continuous laser irradiation to accumulate energy in the DRL and induce the thermal or mechanical changes needed for detachment. The resulting response time---measured from the onset of irradiation on the DRL to chip release---can reach tens or even hundreds of milliseconds, significantly limiting overall throughput.}
This critical limitation is also highlighted in prior studies concerning indirect laser transfer \cite{chen2023laser}. The application of parallel laser pulses represents a potential solution; however, this approach necessitates a complex excimer laser system and a laser mask to enable the patterned distribution of the beam. Consequently, the selectivity of the process is constrained by the design of the mask.

In contrast, the direct laser ablation method provides a more straightforward approach to chip release, where the device is released through the direct ablation of the DRL by a minimum of one laser pulse, which achieves a reaction time of around 1 ns.
To date, only a limited number of investigations \cite{xie2024high, yang2024super} have explored this method for the transfer of micro-LEDs. Furthermore, these studies do not provide a systematic investigation or comprehensive discussion regarding the parameters of the transfer or the performance in precision. Additionally, Xie et al. \cite{xie2024high} utilized a DRL with a specialized design, whereas Yang et al. \cite{yang2024super} did not specify the type of DRL deployed.

In this study, we propose a defocused direct laser ablation transfer (D-DLAT) method for the mass transfer of micro-LEDs. The key idea is the introduction of a defocused laser beam to directly ablate the PI dynamic release layer. This direct ablation generates a highly confined, ultrafast transient shockwave within a timescale of $\sim 1$ ns. Upon ablation, this shockwave actively propels the micro-LEDs toward the receiver, fundamentally altering traditional indirect laser transfer dynamics. Specifically, to the best of our knowledge, we present the first systematic investigation of the direct laser ablation transfer of micro-LEDs. This investigation encompasses laser energy redistribution, optimization of the ablation profile and scanning path, analysis of transfer gap effects, examination of the underlying transfer mechanism, and evaluation of laser-induced damage. Consequently, the D-DLAT method facilitates the transfer of micro-LEDs with dimensions ranging from approximately $600$ \textmu m down to $35$ \textmu m, 
{achieving a minimum placement error of less than $1$ \textmu m},
a transfer gap of $\ge 50$ \textmu m, and a reaction time of $\sim 1$ ns. Furthermore, demonstrations including patterned transfers and the re-pitching of micro-LED arrays are conducted to validate the full selectivity of this method. These results illustrate the substantial potential of the D-DLAT approach, providing a crucial capability for fabricating advanced display panels \cite{Chen2022Mass, linghu2021mass, Wu2020Full-Color, Anwar2022Recent}.

\subsection*{RESULTS} 

\subsubsection*{Defocused direct laser ablation transfer process}

\begin{figure}[] 
	\centering
	\includegraphics[width=0.8\textwidth]{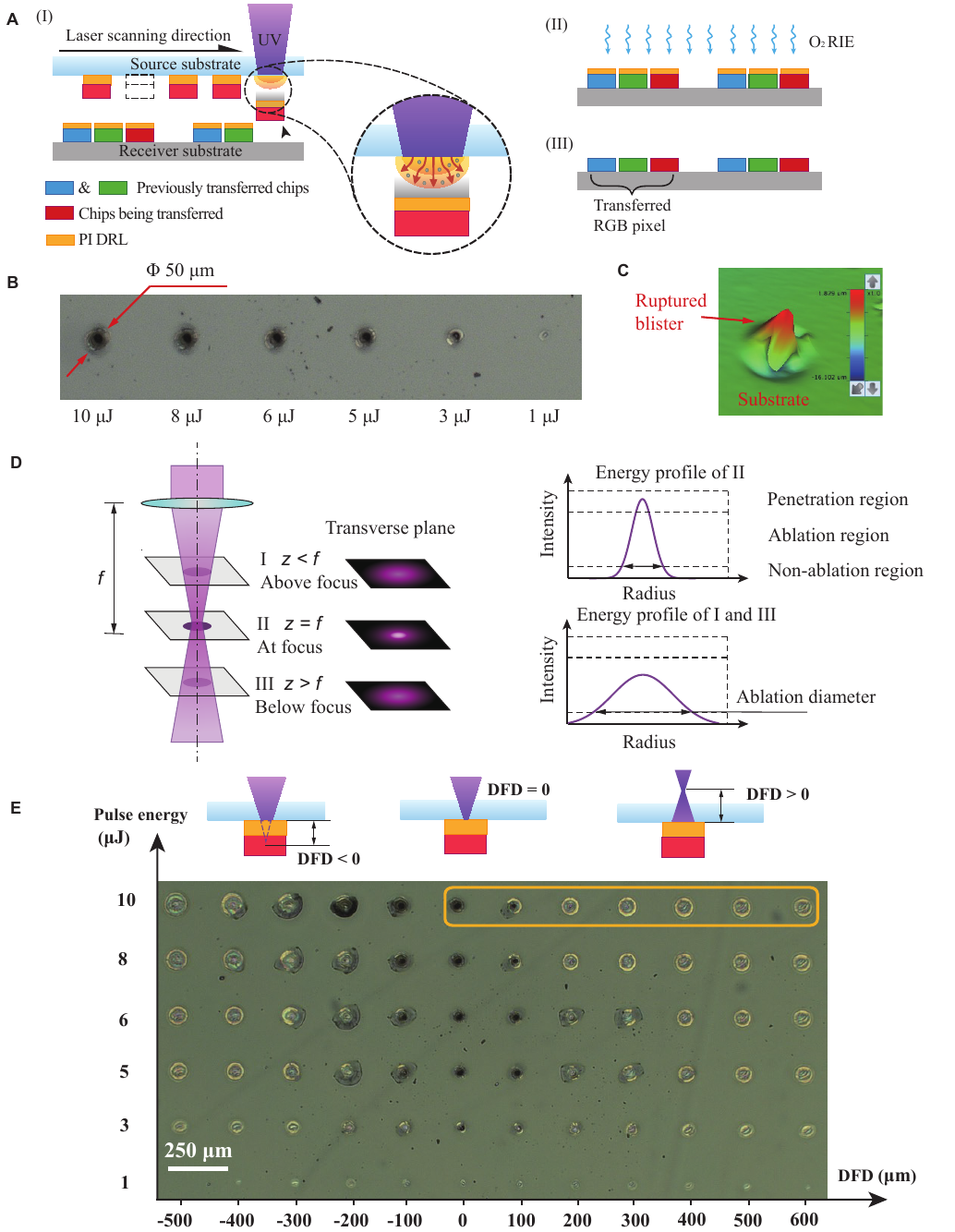} 

	\caption{\textbf{Basic concept of the D-DLAT method.}
	(A) Schematic diagram of the D-DLAT process with illustration on the ablation-induced release mechanism.
	(B) Top-view optical microscopy images of focused PI ablation craters at various pulse energy levels.
	(C) Topography of a focused PI ablation crater under 10 \textmu J pulse energy characterized by WLI. 
	(D) Schematic diagram of the laser energy distribution profiles at the focal plane and under defocused conditions.
	(E) Parameter sweeping results for the ablation profile across various pulse energies and DFD values, with optimal candidate ablations highlighted by the orange rectangle.
	}
	\label{fig:Concept} 
\end{figure}

Figure \ref{fig:Concept}A illustrates a schematic diagram of the entire D-DLAT process. Upon receiving a laser pulse, the PI DRL, with a thickness of 10 \textmu m, situated between the flyer (the micro-LED chip) and the source substrate, undergoes rapid ablation and decomposes into gas and plasma\cite{zhang2022laser}. Once all physical connections to the donor are completely removed via ablation, the flyer is released and descends to the receiver alongside the residual PI layer. Through the precise control of the scanning path and the modulation of the laser beam, exclusively the designated chips are selected and transferred to the receiver. The final position of these chips is determined by the relative movement between the donor and the receiver.
Following the transfer of all chips to the receiver, a reactive ion etching (RIE) process is conducted to remove the residual PI, thereby exposing the emissive surfaces of the micro-LED dies.

To further investigate the laser-induced release manner, the ablation profile on the PI layer is examined. Figure \ref{fig:Concept}B presents top-view optical microscopy images of the ablation craters on the PI layer, which are generated by focused 1 ns, 355 nm UV laser pulses at varying energy levels. Each crater corresponds to irradiation by a single laser pulse. It is clearly observed that an increase in laser energy results in a larger diameter of the crater. Furthermore, several other issues regarding the ablation profile remain.

The first issue is that the diameter of the ablation remains insufficient, with the maximum diameter measuring only approximately 50 \textmu m. For larger chips, which can reach up to 600 \textmu m in length, multiple pulses are required to adequately cover the device and facilitate a successful transfer. An increase in the ablation diameter will reduce the number of pulses required to cover the chip, thereby enhancing the overall transfer efficiency.

The second issue is the non-uniform profile of the ablation, which is indicated by the non-circular creator outline. This lack of uniformity during ablation may result in the release of the chip along a non-vertical trajectory when conducting the single-pulse release of smaller-sized chips, which consequently causes a larger error in placement.

The third issue is the DRL penetration at the center of the crater. The laser pulse generates a blister on the surface of the DRL, which features a penetration at the center. This observation is further verified by the topographic measurements obtained via a white light interferometer (WLI), as illustrated in Figure \ref{fig:Concept}C. It is observed that the ablation generated the profile of a ruptured blister, which indicates that penetration occurred during the process of ablation. This penetration is also evident in the profiles of the ablation at other levels of pulse energy, and this phenomenon may pose potential risks of direct damage to the chips from the gas, plasma, or the laser, which must be avoided.

It is observed that the distribution profile of laser pulse energy may account for the limited ablation diameter and the laser penetration. As illustrated in Figure \ref{fig:Concept}D, the laser energy distribution typically follows a Gaussian profile at the focal point, characterized by a concentrated peak of energy at the center, which decreases significantly with an increase in the distance from the center. Consequently, the energy at the center is sufficiently high to penetrate the PI layer, whereas the energy at the periphery is inadequate to induce the ablation. This discrepancy leads to the observation of an ablation profile of a small and ruptured blister. Upon the application of defocus and the re-evaluation of the energy distribution at the cross-section, the profile becomes more uniform. This uniformity likely results in an non-ruptured intact ablation with a larger diameter.

This assumption is subsequently validated through parameter-sweeping experiments, as illustrated in Figure \ref{fig:Concept}E. A series of experiments on ablation are conducted using various energies of the laser pulse ranging from 1 \textmu J to 10 \textmu J, alongside defocus depth (DFD) ranging from -500 \textmu m to +600 \textmu m, at intervals of 100 \textmu m. To eliminate the risk of the laser direct focusing onto the chip, only the DFD values greater than 0 are considered for subsequent experiments. To maximize the diameter of the ablation, the consideration to the pulse energy are restricted at 10 \textmu J. All candidate profiles for ablation are indicated by the orange rectangle in Figure \ref{fig:Concept}E.

To optimize the ablation parameters, we establish the following three criteria to define a favorable ablation profile:

\begin{enumerate}
	\item A large diameter of the inscribed circle, to ensure a high efficiency of coverage during the ablation;
	\item High circularity, which represents the uniformity and the robustness of the process;
	\item No penetration by the laser, to minimize the potential damage to the chips.
\end{enumerate}

Zoom-in microscope images of the candidate ablations were captured, and computer vision based image processing techniques were employed to extract the contour profile, as shown in Figure \ref{fig:sup_counter}. This process enabled the automatic calculation of circularity and the diameter of the inscribed circle. Whether there's a laser penetration is confirmed together by the optical image and the WLI profile to check whether there's a rupture on the ablation. Table \ref{tab:profile} shows the results of the profile evaluation, with the 300 µm DFD exhibiting the best overall performance. Consequently, the 300 µm DFD and 10 µJ pulse energy is utilized as default for all subsequent transfer experiments.

Magnified images from the microscope of the candidate ablations were captured, and computer vision based image processing techniques were employed to extract the contour profile, as illustrated in Figure \ref{fig:sup_counter}. This process facilitated the automatic calculation of the circularity and the diameter of the inscribed circle. The presence of laser penetration is confirmed through a combination of the optical image and the profile from the WLI to determine the existence of a rupture on the ablation. Table \ref{tab:profile} presents the results of the profile evaluation, wherein the DFD of 300 \textmu m exhibits the optimal overall performance. Consequently, the DFD of 300 \textmu m and the pulse energy at 10 \textmu J are utilized as the default parameters for all subsequent experiments regarding the transfer.

\begin{table} 
	\centering
	\caption{\textbf{Quantitive evaluation of the ablation profile under different DFD values.}
		}
	\label{tab:profile} 
	
	\begin{tabular}{cccccccc} 
		\\
\toprule
DFD/\textmu m& 0    & +100 & +200 & \textbf{+300} & +400 & +500 & +600 \\
\midrule
Diameter/\textmu m & 49.45 & 54.71 & 63.75 & \textbf{68.01} & 67.85 & 64.45 & 63.57 \\

Circularity/\% & 96.78 & 98.48 & 99.24 & \textbf{99.29} & 99.05 & 99.09 & 98.81 \\

Rupture?  & Yes   & Yes  & No   & \textbf{No}   & No   & No   & No \\
\bottomrule
	\end{tabular}
\end{table}

\subsubsection*{Multi-pulse laser transfer}

\begin{figure}[!htb] 
	\centering
	\includegraphics[width=\textwidth]{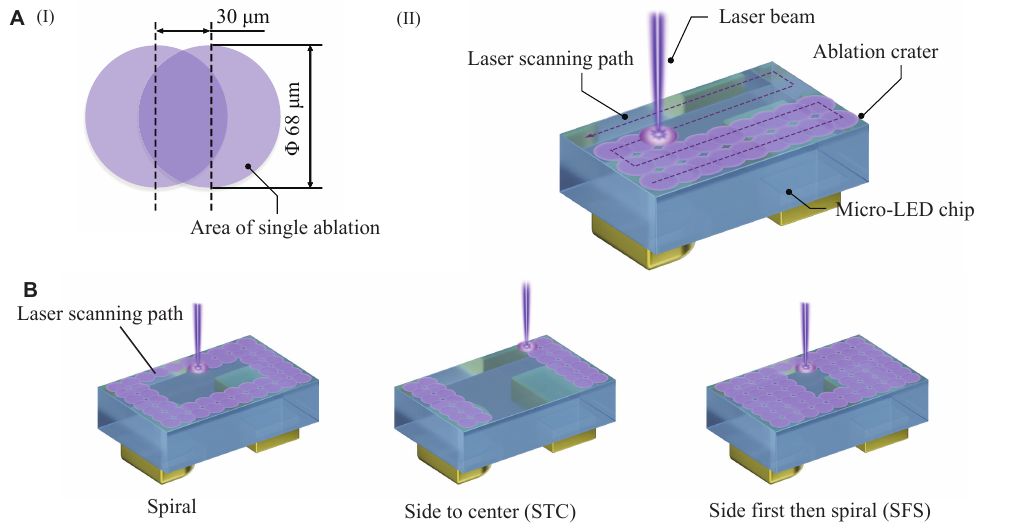} 

	\caption{\textbf{Schematic of the multi-pulse laser transfer process.}
		(A) Schematic diagrams of: (I) the ablation overlap at a 30 µm spacing, and (II) the line-by-line laser scanning pattern.
		(B) Schematic diagram of the three designed laser scanning patterns for multi-pulse transfer.
	}
	\label{fig:Multi} 
\end{figure}

For the chips (in our case, 570 $\times$ 270 $\times$ 400 \textmu m and 260 $\times$ 150 $\times$ 120 \textmu m), which are significantly larger than the dimension of laser ablation(68 \textmu m in diameter), a single pulse is insufficient to cover the entire surface of the chip and trigger a transfer. Consequently, an array of pulses is necessary to accomplish the transfer of the chip. Initially, a line-by-line laser scanning process was attempted on the chips with dimensions of 570 $\times$ 270 $\times$ 400 \textmu m, wherein the spacing between the pulses was set to 30 \textmu m, as illustrated in Figure \ref{fig:Multi}A. The chips remained untransferred until the execution of the final several ablations at the corners. However, all the transferred chips exhibited significant displacement, often exiting the field of view and exceeding the measurement limits. It is hypothesized that this displacement may be attributed to the torque induced by the gravitational force acting upon the mass of the chip, as demonstrated in Figure \ref{fig:sup_gravity}.

\begin{figure}[!htb] 
	\centering
	\includegraphics[width=\textwidth]{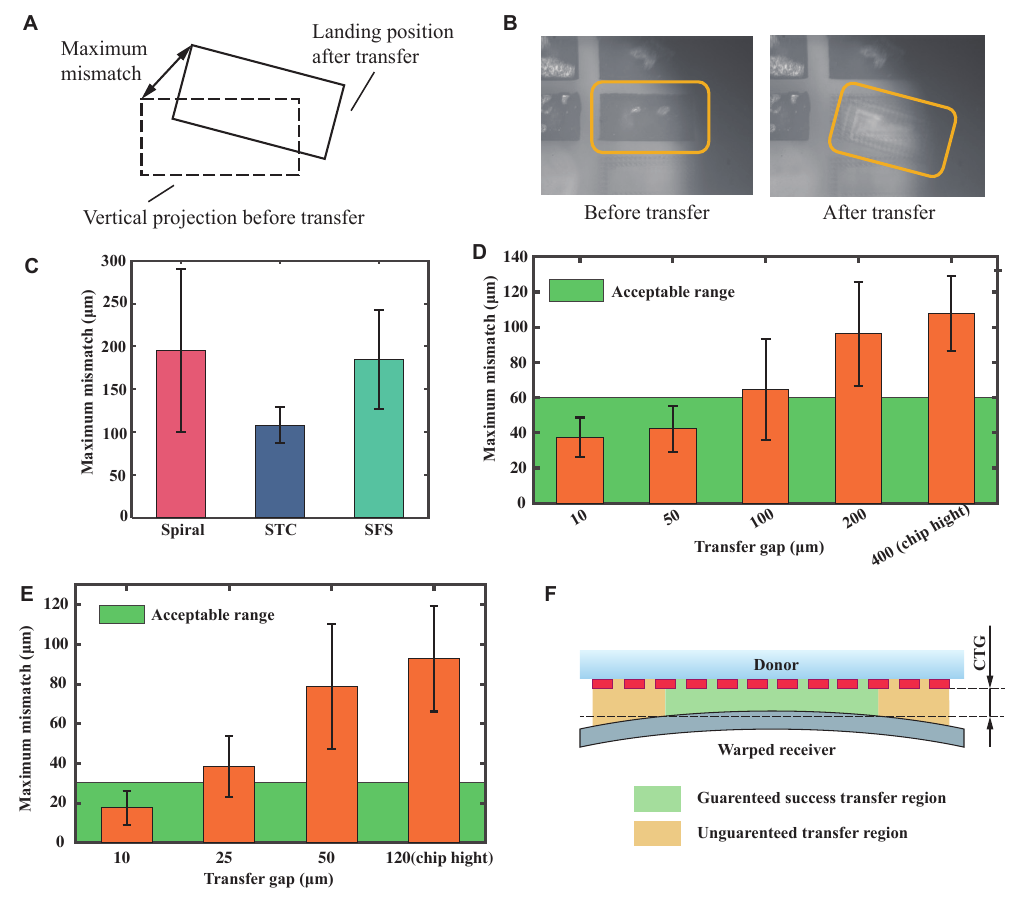} 

	\caption{\textbf{Precision results of the multi-pulse laser transfer.}
		(A) Schematic illustration of the mismatch calculation for the transferred chips.
		(B) Optical microscopy images of the chips before and after transfer, with the transferred chip highlighted by the orange rectangle.
		(C) Transfer precision achieved using the three designated laser scanning patterns.
		(D) Transfer precision of the side-to-center scanning pattern across varying transfer gaps for chips measuring 570 × 270 × 400 µm.
		(E) Transfer precision of the side-to-center scanning pattern across varying transfer gaps for chips measuring 260 × 150 × 120 µm.
		(F) Schematic diagram of the wafer flatness issue and its consequent impact on transfer throughput.}
	
	\label{fig:MultiResult} 
\end{figure}

To address this issue, three laser scanning patterns are designed and evaluated to ensure that the final laser ablation occurs near the center of the chip, as illustrated in Figure \ref{fig:Multi}B. Chips measuring $570 \times 270 \times 400$ \textmu m are transferred across a $400$ \textmu m gap using these patterns, and the resulting transfer mismatches are quantified, as presented in Figure \ref{fig:MultiResult}A, B, and C. The scanning pattern progressing from the side to center achieves the highest precision and will be utilized in subsequent experiments.

To further improve precision and limit the maximum displacement to 10\% of the chip dimension (57 \textmu m), the transfer gap is decreased, with the precision results detailed in Figure \ref{fig:MultiResult}D. It is evident that reducing the transfer gap minimizes mismatches, thereby enhancing precision. When the gap is reduced to less than 50 \textmu m---approximately one-eighth of the chip height---the mismatch is deemed acceptable, falling below 10\% of the chip length. Similar gap-performance experiments conducted on chips measuring $260 \times 150 \times 120$ \textmu m yield consistent findings: the transfer mismatch becomes acceptable only when the gap is reduced to less than one-eighth of the chip height (Figure \ref{fig:MultiResult}E).

The critical transfer gap (CTG) is defined as the upper limit of the transfer gap that ensures a successful chip transfer. The CTG serves as a crucial performance metric for two primary reasons. First, it addresses the issue of wafer flatness. Across a standard 6-inch wafer, the height variation between the center and the edge can range from tens to hundreds of micrometers. In industrial settings where wafer-to-wafer transfer is prevalent, a larger CTG allows broader areas of the wafer to remain within the successful transfer threshold. This facilitates the transfer of a greater number of chips in a single batch, thereby enhancing overall throughput, as illustrated in Figure \ref{fig:MultiResult}F. Second, a large CTG is necessary for the full-color integration of micro-LED displays. Integrating three colors of micro-LED chips onto a receiving driving backplane requires at least three separate transfer processes, as depicted in Figure \ref{fig:Concept}A. To prevent physical interference between the chips already assembled on the receiver and those remaining on the donor substrate awaiting transfer, a CTG exceeding the chip thickness is essential. Consequently, a larger CTG provides greater gap tolerance and is highly desirable.

{The large-sized chips discussed in this section are rarely used directly in full-color displays; instead, they typically serve as white backlight sources for display panels\cite{wu2018mini}. In this case, the single-color backlight configuration requires only one chip-transfer sequence, rather than the three required for full-color integration, so physical interference is not a concern.}
 Therefore, a CTG smaller than the chip thickness remains acceptable. However, for small chips intended to directly form RGB pixel arrays, a CTG exceeding the chip thickness is strictly essential.

\subsubsection*{Single-pulse laser transfer}\label{sec:Single}

\begin{figure}[!htb] 
	\centering
	\includegraphics[width=\textwidth]{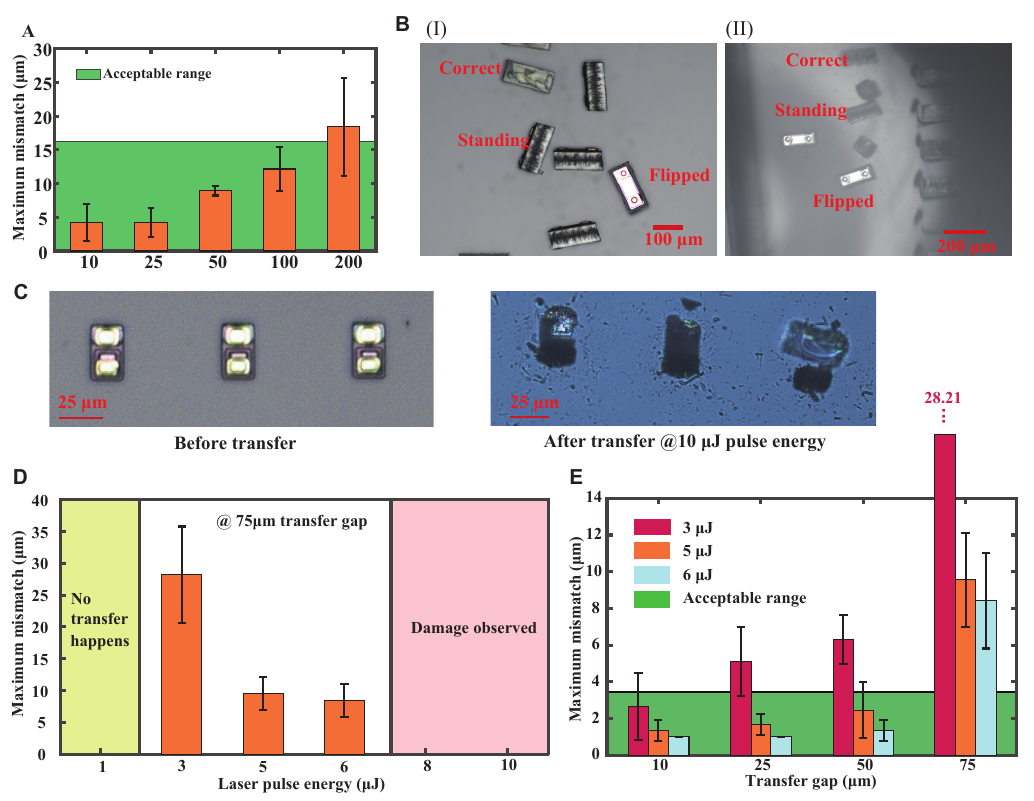} 

	\caption{\textbf{Precision results of the single-pulse laser transfer.}
	(A) Transfer precision of the single-pulse laser transfer for chips measuring 160 × 70 × 70 µm across varying transfer gaps.
	(B)	Optical microscopy images of chips measuring 160 × 70 × 70 µm transferred without laser defocusing, captured by: (I) a top-view microscope, and (II) the \textit{in-situ} camera integrated into the laser system.
	(C) Optical microscopy images of chips measuring 35 × 20 × 5 µm transferred using a 10 µJ pulse energy, acquired via a bottom-view microscope.
	(D) Transfer performance of chips measuring 35 × 20 × 5 µm across various pulse energy levels.
	(E) Transfer precision for chips measuring 35 × 20 × 5 µm under different pulse energy levels and transfer gaps.	}
	\label{fig:Single} 
\end{figure}

When the chip dimensions are further reduced to $160 \times 70 \times 70$ \textmu m, the chip area approaches that of the laser ablation spot, which possesses a diameter of $68$ \textmu m. Under these conditions, a single laser pulse is sufficient to trigger the chip transfer. The precision test results are presented in Figure \ref{fig:Single}A. Notably, the CTG exceeds $100$ \textmu m, a value significantly larger than the chip thickness. The underlying mechanism responsible for this extended CTG is detailed in the subsequent mechanism study section.

Given the highly favorable precision and mismatch performance, a comparative study involving the transfer of identical chips without laser defocusing is conducted to demonstrate the improvements achieved by this technique. In this comparative experiment, all chips are transferred across a $100$ \textmu m gap. In the batch processed without laser defocusing, $46.2$\% of the chips landed on their sides, whereas $15.4$\% flipped upside down during transfer with their electrodes facing upwards, as illustrated in Figure \ref{fig:Single}B. Furthermore, all samples in the non-defocused batch, including those correctly oriented without flipping, exhibit significant transfer mismatches. In contrast, chips transferred with a $300$ \textmu m defocus (the parameter applied in all other transfer experiments) are transferred with $100$\% accuracy, demonstrating a maximum mismatch of $<16$ \textmu m. This comparison clearly demonstrates the substantial enhancement in precision enabled by laser defocusing.

When the chip dimensions are further reduced to $35 \times 20 \times 5$ \textmu m, the device area becomes significantly smaller than that of the laser ablation spot. An initial test is conducted utilizing a default laser energy of $10$ \textmu J and a $75$ \textmu m transfer gap. Three chips are transferred, exhibiting an average mismatch of $2.60$ \textmu m. However, subsequent microscopic inspection reveals that all three chips are severely fractured, as shown in Figure \ref{fig:Single}C. This indicates that critical structural damage occurs during the transfer process, rendering the current parameters unacceptable.

To prevent structural damage and ensure chip functionality, the laser pulse energy is reduced, and the precision tests are repeated, as depicted in Figure \ref{fig:Single}D. When the pulse energy is decreased to $6$ \textmu J, chip fracturing is completely eliminated. Further reducing the energy to $1$ \textmu J results in insufficient power to generate an ablation spot strong enough to initiate the transfer. For the $3$ \textmu J, $5$ \textmu J, and $6$ \textmu J energy levels, which guarantee damage-free transfers, experiments across various gaps are conducted to evaluate transfer precision, as illustrated in Figure \ref{fig:Single}E. At a given transfer gap, higher laser energies consistently correlate with significantly reduced positional mismatches and higher precision. Furthermore, the CTG for higher-energy transfers is considerably larger than that for lower-energy transfers. Specifically, the CTG for the $6$ \textmu J transfer exceeds $50$ \textmu m, measuring more than ten times the chip height. 
{With the image scale used for displacement measurement calibrated to be $0.97$ \textmu m per pixel, all displacements below this threshold are conservatively recorded as $0.97$ \textmu m. }
Consequently, the actual transfer precision achieved with a $6$ \textmu J energy and a $10$ \textmu m gap is likely superior to the data presented in Figure \ref{fig:Single}E.

\subsubsection*{Discussion on the transfer mechanism}

The transfer precision of the D-DLAT method for large chips requiring multi-pulse transfers differs significantly from that of small chips relying on single-pulse transfers. This discrepancy is attributed to fundamental differences in the underlying transfer mechanisms, which are detailed in this section.

For large chips, the majority are not transferred until the final laser pulse is applied to ablate the remaining PI between the chip and the donor glass slide. The remaining chips detach spontaneously prior to the final ablation, while a minimal amount of PI remains. This observation suggests that, for large chips, gravitational force contributes significantly to the process and serves as the primary factor facilitating the transfer. Consequently, the transfer mechanism for these chips closely resembles a free-falling process.

\begin{figure}[!htb] 
	\centering
	\includegraphics[width=\textwidth]{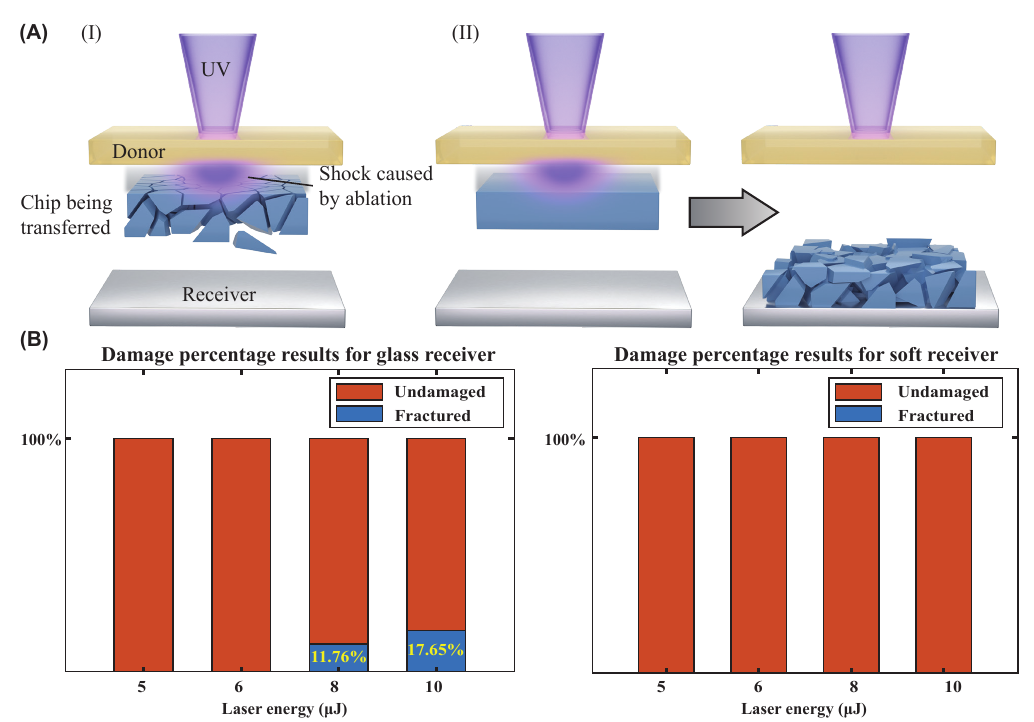} 

	\caption{\textbf{Investigation of the single-pulse laser transfer mechanism via damage evaluation.}
	(A) Schematic diagram of the two hypotheses regarding the damage mechanism. Hypothesis (I): Chips incur damage from laser-induced shock during the ablation process. Hypothesis (II): Chips sustain damage from the impact force upon landing on the receiver.
	(B) Damage evaluation results for chips measuring 35 × 20 × 5 µm across various pulse energy levels and receiver materials.
	}
	\label{fig:Damage} 
\end{figure}

The transfer mechanism for small chips is incidentally revealed during the damage evaluation process. As discussed in the single-pulse transfer section, applying a $10$ \textmu J laser pulse results in structural damage to the transferred chip. To explain this phenomenon, two hypotheses are proposed: the damage occurs either during the laser ablation process, or upon impact when landing on the receiver after the flight. Figure \ref{fig:Damage}A illustrates the schematics of these two hypotheses. Under the first hypothesis, the laser ablation generates a substantial volume of plasma and gas, potentially inducing direct shock damage to the chip. Under the second hypothesis, the chip is propelled across the gap and sustains damage upon landing on the receiver owing to the high impact velocity and the corresponding impact force.

To verify these hypotheses, comparative experiments are conducted utilizing different receiver materials. Chips measuring $35 \times 20 \times 5$ \textmu m are transferred using a $10$ \textmu J laser pulse across a $75$ \textmu m gap, employing a glass slide and soft packing foam as the receivers, respectively. The glass slide exhibits a Young's modulus of $>50$ GPa, whereas the foam material demonstrates a modulus of merely $1$ to $10$ kPa. Consequently, it is reasonable to assume that the foam is sufficiently soft compared with the glass slide, rendering the impact between the chip and the foam too weak to cause structural damage. Therefore, any chip damage observed in the foam receiver group can be exclusively attributed to the laser shock.

Laser pulses of $10$ \textmu J, $8$ \textmu J, and $6$ \textmu J are applied in the experiments, and the corresponding results are detailed in Figure \ref{fig:Damage}B. In the glass receiver group, the observed chip fracturing rates are $17.65$\%, $11.76$\%, and $0$\%, respectively. Conversely, the foam receiver group exhibits zero fracturing across all transferred chips. These findings suggest that the structural damage observed on the chips is induced by the collision between the chip and the receiver. Furthermore, an increase in laser energy consistently correlates with a higher percentage of damaged chips. This trend implies the occurrence of more severe impacts, which are a direct consequence of higher flight velocities.

From these results, it can be concluded that in the scenario of single-pulse transfer, an increase in laser energy corresponds to a higher flight velocity of the chips. This conclusion also indicates that single-pulse transferred chips are actively propelled towards the receiver during laser DRL ablation, whereas multi-pulse transferred chips follow a nearly free-falling mechanism lead by gravity.

Consequently, the flight time of actively propelled chips (single-pulse transfer) is significantly shorter than that of freely falling chips (multi-pulse transfer). This reduction in flight time minimizes potential disturbances during flight and contributes to the observed improvements in transfer precision and CTG. Specifically, the CTG for single-pulse transfers can exceed ten times the chip height, whereas the CTG for multi-pulse transfers remains less than one-eighth of the chip height. Furthermore, during single-pulse transfers, utilizing higher laser energies yields greater flight velocities and enhanced process robustness, directly resulting in superior transfer precision and a larger CTG. Therefore, applying higher laser energy in single-pulse transfers is highly desirable, provided that the energy remains below the structural damage threshold of the chips.

\subsubsection*{Selective transfer demonstration}

The full selectivity of the D-DLAT method is demonstrated in selective transfer experiments. By precisely controlling the relative motion between the donor and the receiver, in conjunction with the laser scanning pattern, both re-pitching and patterned transfer are effectively achieved.

\begin{figure}[!htb] 
	\centering
	\includegraphics[width=0.85\textwidth]{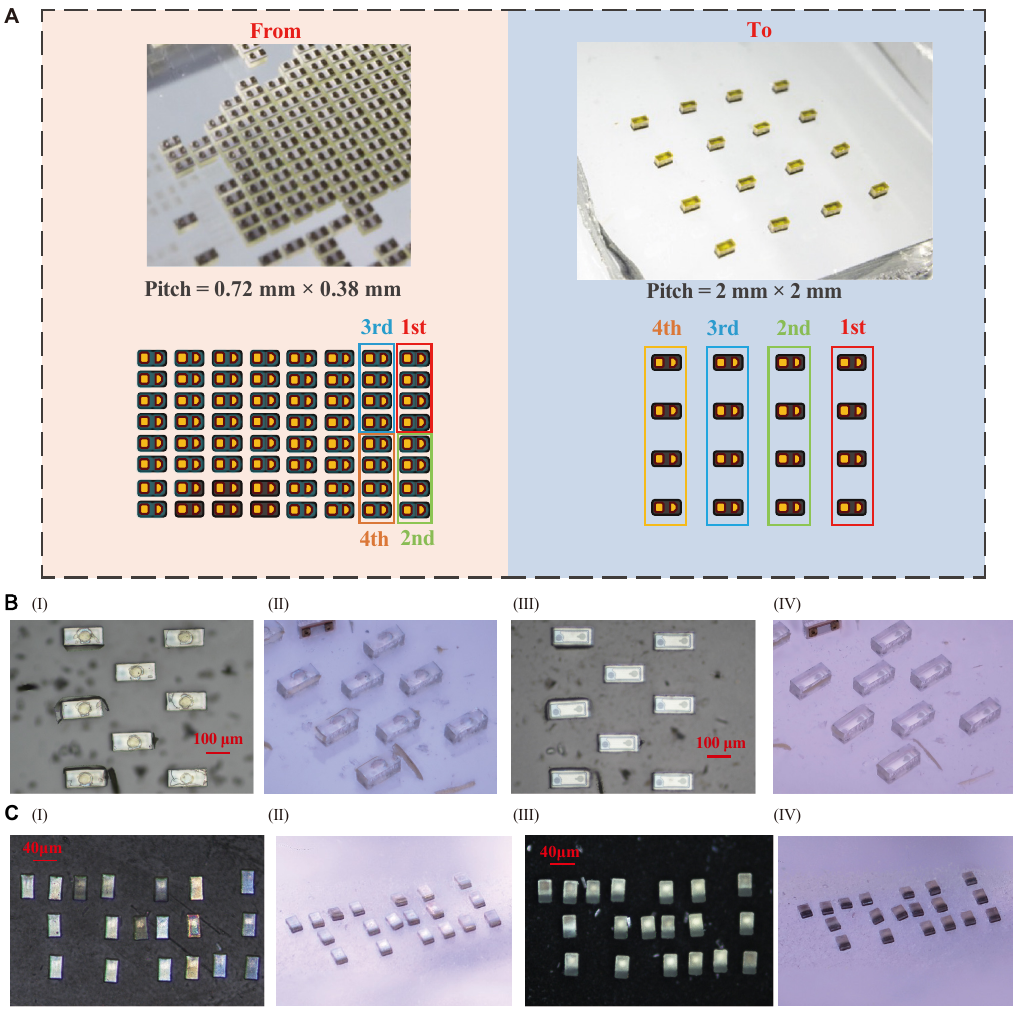} 

	\caption{\textbf{Demonstrations of selective transfer capabilities.}
	(A) Re-pitching of chips measuring 570 × 270 × 400 µm from an original pitch of 0.72 ×  0.38 mm on the donor to a new pitch of 2 × 2 mm on the receiver.
	(B) Re-pitching and cross-shaped patterned single-pulse transfer of chips measuring 160 × 70 × 70 µm. (I) Optical microscopy image and (II) macroscopic photograph of the transferred chips on the receiver prior to RIE. (III) Optical microscopy image and (IV) macroscopic photograph of the transferred chips on the receiver post-RIE. 
	(C) Re-pitching and THU''-shaped patterned transfer of chips measuring 35 × 20 × 5 µm. (I) Optical microscopy image and (II) macroscopic photograph of the transferred chips on the receiver prior to RIE. (III) Optical microscopy image and (IV) macroscopic photograph of the transferred chips on the receiver post-RIE.
	}
	\label{fig:Selective} 
\end{figure}

Figure \ref{fig:Selective}A illustrates the re-pitching of chips measuring $570 \times 270 \times 400$ \textmu m at a transfer gap of $50$ \textmu m. By precisely controlling the relative motion, the chips are transferred from an original pitch of $0.72 \times 0.38$ mm on the donor to a new pitch of $2 \times 2$ mm on the receiver, representing a common scenario for white backlight sources or ``Micro-LED in package'' (MIP) solutions in display panels\cite{Fan2023532, Huang2025GlassBased}. 

For smaller-sized chips with a CTG exceeding the chip thickness, no physical interference occurs between the donor and the assembled chips on the receiver; consequently, simultaneous re-pitching and patterned transfer can be achieved. Figure \ref{fig:Selective}B illustrates the re-pitching and cross-shaped patterned single-pulse transfer of chips measuring $160 \times 70 \times 70$ \textmu m at a transfer gap of $100$ \textmu m. Meanwhile, Figure \ref{fig:Selective}C depicts the re-pitching and ``THU''-shaped patterned transfer of chips measuring $35 \times 20 \times 5$ \textmu m at a transfer gap of $50$ \textmu m. These results demonstrate the full selectivity of the D-DLAT method, a capability crucial for fabricating advanced full-color display panels.

\subsubsection*{Damage evaluation}

Although the macroscopic structure of the chips is maintained following the D-DLAT process, further characterization is necessary to evaluate the photoelectric performance and assess potential internal damage. The reference samples consist of bare micro-LED dies directly detached from the wafer, whereas the D-DLAT samples comprise chips subjected to the complete donor preparation procedure, the laser ablation transfer, and the subsequent RIE residue cleaning. This comparison aims to quantify any overall performance degradation induced by the entire D-DLAT method.

\begin{figure}[!htb]
	\centering
	\includegraphics[width=\textwidth]{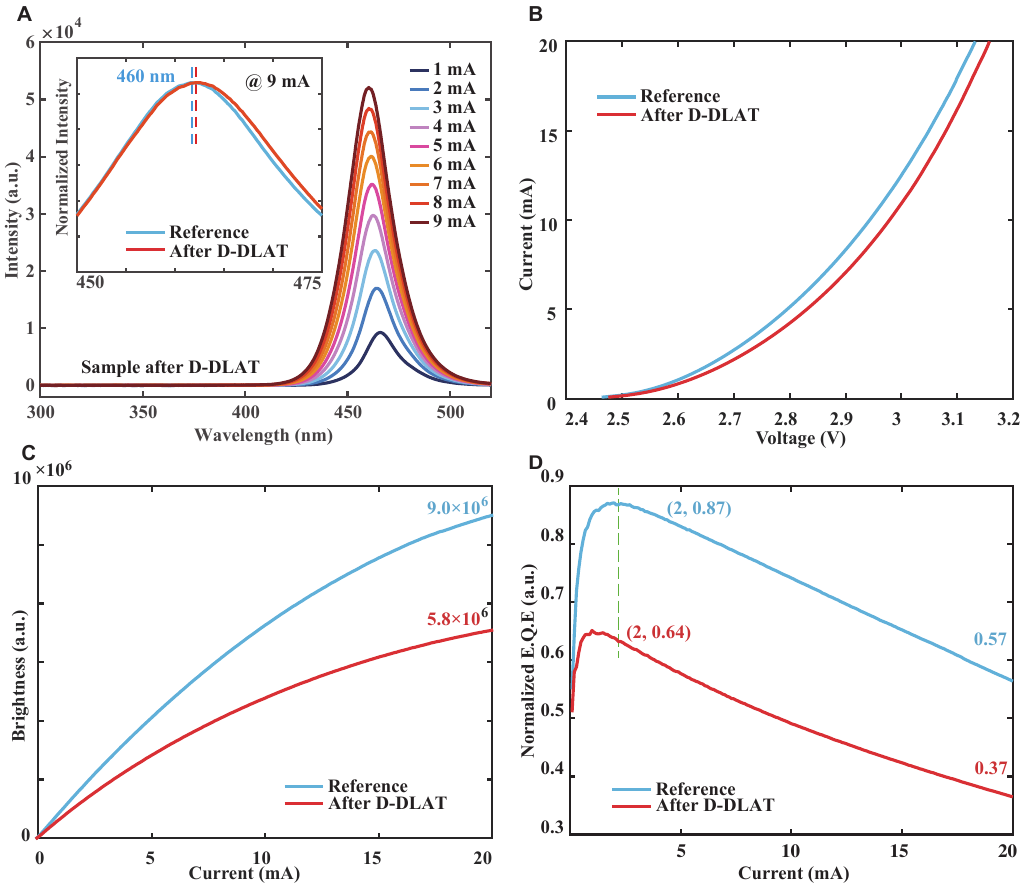} 

	\caption{\textbf{Damage evaluation results of the D-DLAT process.}
	(A) EL spectra of the reference and D-DLAT samples under an injection current of 9 mA..
	(B) I--V characteristics of the reference and D-DLAT samples.
	(C) Current-brightness curves of the reference and D-DLAT samples.
	(D) EQE curves of the reference and D-DLAT samples.
	}
	\label{fig:Spec} 
\end{figure}

Figure \ref{fig:Spec}A shows the electroluminescence (EL) spectra of the reference and D-DLAT samples under an injection current of 9 mA. The EL emmision peak of the D-DLAT sample is slightly red-shifted by 1 nm, which remains within the commercial tolerance (\textpm \ 4 nm) and is not significant. Figure \ref{fig:Spec}B shows the I–V characteristics of the two batches. A slight rightward shift is observed for the D-DLAT sample compared with the reference sample, indicating a modest increase in parasitic resistance and turn-on voltage induced by the whole transfer process. Figure \ref{fig:Spec}C and D shows the current-brightness curves and the external quantum effeciency (EQE) curves, respectively, where the brightness and EQE of the transferred chips shows a maximum decrease of $\approx$ 35\%, but remains the same trend and order of magnitude as the reference chips.  This reduction remains within a reasonable range compared with previously reported results, indicating that the D-DLAT process imposes an acceptable influence on the overall performance of the micro-LEDs.

Figure \ref{fig:Spec}A illustrates the electroluminescence (EL) spectra of the reference and D-DLAT samples under an injection current of $9$ mA. The EL emission peak of the D-DLAT sample exhibits a slight red-shift of $1$ nm, which remains within the commercial tolerance ($\pm 4$ nm) and is considered negligible\cite{sun202420}. Figure \ref{fig:Spec}B depicts the I--V characteristics of the two batches. A slight rightward shift is observed for the D-DLAT samples compared with the reference samples, indicating a modest increase in parasitic resistance and turn-on voltage induced by the entire transfer process. Figures \ref{fig:Spec}C and D display the current-brightness and external quantum efficiency (EQE) curves, respectively. The brightness and EQE of the transferred chips show a maximum decrease of $\approx 35$\%, but maintain the same trends and order of magnitude as those of the reference chips. This reduction remains within a reasonable range compared with previously reported results, indicating that the D-DLAT process exerts an acceptable impact on the overall performance of the micro-LEDs\cite{qi2021848, jang2022auxetic}.

\subsection*{Conclusion and Future Work}

{In conclusion, an ablation-driven direct laser chip transfer method for the mass transfer of micro-LEDs, termed D-DLAT, is reported and systematically investigated for the first time. This method demonstrates the capability to transfer micro-LED chips with high precision, large gap tolerance, and full selectivity. 
Direct laser ablation of the DRL propels micro-sized chips toward the receiver, enabling high-precision transfer across large gaps.
Furthermore, the defocused laser energy profile facilitates mild ablation and prevents laser penetration through the DRL. This mitigates potential laser-induced damage to the chips while simultaneously improving transfer precision and the overall success rate.
}

{In the single-pulse transfer mode, the D-DLAT method actively propels the chips towards the receiver, achieving a minimum transferable chip size of 35 \textmu m, a minimum placement error of $<$1~\textmu m, and a maximum transfer gap of $\ge$50 \textmu m, which exceeds ten times the chip height. Furthermore, this method is compatible with a multi-pulse transfer mode for larger chips governed by a release followed by passively descend mechanism, successfully demonstrating the transfer of chips with lengths up to 570 \textmu m in this study. 
Patterned transfer demonstrations further show the method's full selectivity, a capability crucial for fabricating advanced display panels. A detailed performance comparison with alternative laser transfer methods is presented in Table \ref{tab:comparison}.
}

\begin{table} 
	\centering
	\caption{\textbf{Performance comparison of laser-assisted micro-LED transfer methods.}
		Comparison of minimum chip size, placement precision, maximum transfer gap, and reaction time for existing laser transfer techniques and the proposed D-DLAT method.}
	\label{tab:comparison} 
	
\begin{tabular}{ccccc}
\toprule
Transfer method & Min chip size & Best precision & Max transfer gap & Min reaction time \\
\midrule
Blister           & 35 \textmu m \cite{huang2025high} & 2 \textmu m \cite{huang2025high}  & 25 \textmu m \cite{huang2025high} & 20 ms \cite{li2024laser} \\
Thermal           & 100 \textmu m \cite{gai2026gradient} & 5 \textmu m \cite{gai2026gradient}  & —     & 10 ms \cite{luo2024switchable} \\
Adhesion reduction & 150 \textmu m \cite{chen2023laser} & 4 \textmu m \cite{chen2023laser}  & 65 \textmu m \cite{chen2023laser} & 200 ms \cite{chen2023laser} \\
D-DLAT & 35 \textmu m & $<$ 1 \textmu m & 50 \textmu m &  1 ns \\
\bottomrule
\end{tabular}
\end{table}

{Future work will explore beam shaping to enlarge the ablation area for single-pulse transfer of larger chips. High-speed imaging will be used to quantify chip flight dynamics and refine the transfer model. Electrical integration of transferred chips with driving circuits will enable evaluation of the method in display arrays. }

\subsection*{MATERIALS AND METHODS}
\subsubsection*{Source substrate fabrication}

{As illustrated in Figure \ref{fig:sup_source}, the source substrate is fabricated following a heterogenous integration process as discribed in our previous study\cite{cao2026heterogeneous}.} It starts by first spin-coating a 10 \textmu m-thick PI layer onto the glass substrate, followed by soft-baking the PI layer at 100 \textdegree C for 30 minutes. Subsequently, the micro-LED chips are attached to the PI layer, which is then baked at 150 \textdegree C for 30 minutes and ultimately hard-cured at 200 \textdegree C for 1 hour. Following the curing process, the original encapsulation layer of the micro-LED chips is removed. An RIE treatment (Oxford Instruments PlasmaPro 80 RIE) is then applied to etch off the residual PI between the chips, {thereby removing the PI connections and forming separate PI DRLs directly beneath them.} The chips adhere to the PI layer exclusively via non-covalent intermolecular interactions without any additional adhesive layer, which is crucial for the laser ablation and subsequent transfer processes.

\subsubsection*{Motion and laser control system setup}

Figure \ref{fig:sup_machine} illustrates the LIFT system setup. Two digital-controlled positioning stages enable precise control of the relative motion between the donor and the receiver along the X, Y, and Z axes with a minimum resolution of 1 \textmu m. A pulsed laser system (355 nm wavelength, 1 ns pulse width) generates individual laser pulses with energies ranging from 1 \textmu J to 10 \textmu J. A coaxial microscope system facilitates \textit{in-situ} observation of the laser ablation and the resulting transfer displacement at an resolution of 0.97 \textmu m, thereby aiding in the evaluation of transfer precision and the optimization of process parameters. Furthermore, a cross-correlation image processing method is employed to precisely quantify the placement mismatch of the $35 \times 20 \times 5$ \textmu m chips, as detailed in Note S1 and Figure \ref{fig:sup_corr}.

\subsubsection*{Photoelectric performance evaluation}

Electrical connections to the chips are established using a probe station equipped with micromanipulator probes. All micro-LED samples are driven by a source meter (Keithley 2400). The I--V characteristics are evaluated by sweeping the injection current via the source meter, whereas the brightness and EL spectra are recorded using a spectrometer (Ocean Optics) coupled with an integrating sphere.

\clearpage

\clearpage 

%
\bibliography{science_template} 
\bibliographystyle{sciencemag}

%
%
%
%
%
%


\section*{Acknowledgments}
The authors gratefully acknowledge Dr. Arnaud Bertsch, and Dr. Zhiwei Yang (Microsystems Laboratory, EPFL) for their insightful discussions and inspiration regarding the characterization and transfer experiments.

\paragraph*{Funding:}
The authors sincerely acknowledge the financial support from the National Natural Science Foundation of China (Grant Nos. 52275564, U24A20109).
\paragraph*{Author contributions:}
Conceptualization: Y.C. Methodology: Y.C., Y.W. Visualization: Y.C., Y.W.  Supervision: J.B., Z.Z. Writing – original draft: Y.C. Writing – review \& editing: Y.C., Y.W., J.B., Z.Z.
\paragraph*{Competing interests:}
There are no competing interests to declare.
\paragraph*{Data and materials availability:}
All data needed to evaluate the conclusions in the paper are present in the paper and/or the Supplementary Materials.
\subsection*{Supplementary materials}

Note S1\\
Figures S1 to S5\\


\newpage


\renewcommand{\thefigure}{S\arabic{figure}}
\renewcommand{\thetable}{S\arabic{table}}
\renewcommand{\theequation}{S\arabic{equation}}
\renewcommand{\thepage}{S\arabic{page}}
\setcounter{figure}{0}
\setcounter{table}{0}
\setcounter{equation}{0}
\setcounter{page}{1} 


\begin{center}
\section*{Supplementary Materials for\\ \scititle}

Yuxuan~Cao, 
Yulong~Wang, 
Jürgen Brugger$^\ast$,
Zhen~Zhang$^\ast$
\\ 
\small$^\ast$Corresponding author. Email: example@mail.com\\
\end{center}

\subsubsection*{This PDF file includes:}
Note S1\\
Figures S1 to S5\\



\subsubsection*{Note S1: Calculation of mismatch using cross-correlation image processing method}

\begin{figure} 
	\centering
	\includegraphics[width=0.8\textwidth]{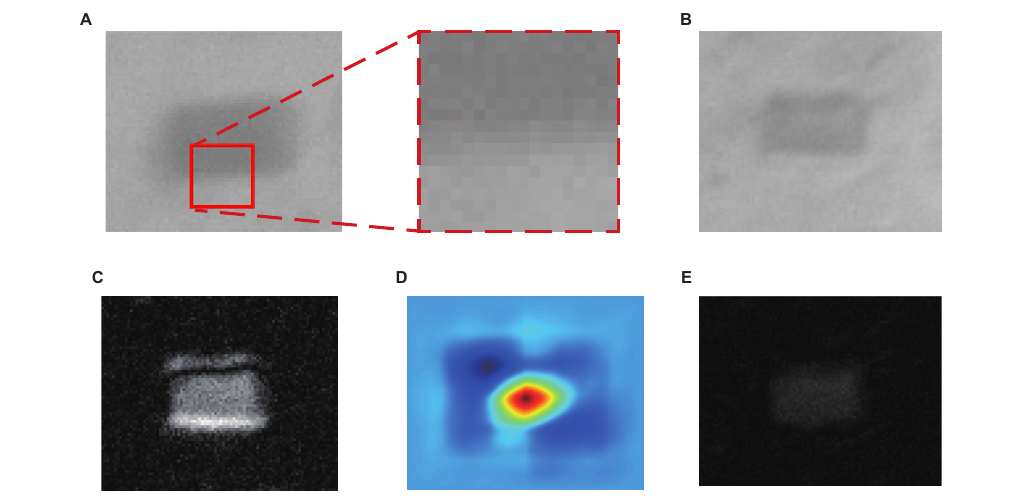} 

	\caption{\textbf{Cross-correlation image processing method for the precise calculation of the mismatch of 35 × 20 × 5 µm chips.}
	 (A) Optical image of the chip before transfer with a zoomed-in view of the edge of the chip. (B) Optical image of the chip after transfer. (C) MSE image of the two images in (A) and (B), while the brighter regions indicate areas of high difference. (D) Image showing the normalized cross-correlation value between (A) and (B), with the maximum value found at (0, -4), indicating the translational mismatch of the chip is 4 pixels, which corresponds to 3.88 \textmu m. (E) MES image after compensating the translational mismatch.
	}
	
	\label{fig:sup_corr} 
\end{figure}

Due to the limited optical resolution of the imaging system, the chip edges lack distinct clarity in the captured images, thereby complicating the precise identification of the chip outline and the subsequent quantification of the placement mismatch. Figure \ref{fig:sup_corr}A illustrates the blurred edge of a 35 × 20 × 5 µm chip, exhibiting a boundary transition region of approximately 7 µm.

To precisely quantify the placement mismatch, specialized image processing techniques are implemented. Although deep learning-based detection frameworks (e.g., YOLO) exhibit exceptional performance in generalized computer vision tasks, their direct application to precise chip displacement quantification presents inherent limitations. First, these data-driven approaches require extensive labeled datasets for training and inherently rely on statistical inference. Consequently, their probabilistic outputs are susceptible to unpredictable generalization errors. Second, they fail to utilize the strong prior that "the images before and after are acquired from the same chip"—that is, the two images are almost homogeneous, with only translational and rotational transformations. This leads to the loss of information such as the microscopic texture of the chip surface.

Consequently, this study employs a cross-correlation algorithm to precisely quantify the translational displacement. Just as its application in determining the time delay between similar one-dimensional signals, the cross-correlation function effectively extracts the spatial displacement vector between homogeneous two-dimensional images.

A grayscale image of size $M \times N$ is typically represented as a real-valued matrix $\mathbf{X} \in \mathbb{R}^{M \times N}$, where each entry $x_{ij}$ denotes the grayscale intensity of the pixel at row $i$ and column $j$. Let the grayscale images acquired before and after chip transfer be denoted by $\mathbf{X} \in \mathbb{R}^{M \times N}$ and $\mathbf{Y} \in \mathbb{R}^{M \times N}$, respectively, where $x_{ij}$ and $y_{ij}$ represent the grayscale values at row $i$ and column $j$ in the two images. Let $(u,v)$ denote the relative displacement in the row and column directions, respectively. For a given displacement $(u,v)$, the valid overlap region between the two images is defined as
\begin{equation}
\Omega_{u,v} =
\left\{
(i,j) \,\middle|\,
\begin{array}{l}
1 \leq i \leq M,\quad 1 \leq j \leq N,\\
1 \leq i+u \leq M,\quad 1 \leq j+v \leq N
\end{array}
\right\}.
\end{equation}

Within this overlap region, the two-dimensional cross-correlation function is
\begin{equation}
R_{X,Y}(u,v) =
\sum_{(i,j)\in\Omega_{u,v}} x_{ij}y_{i+u,j+v}.
\end{equation}

Thus, $R_{X,Y}(u,v)$ describes the correlation distribution as a function of the two-dimensional displacement $(u,v)$. As $u$ and $v$ vary, the two images are compared at different relative positions. The cross-correlation reaches its maximum when the chip outline, surface texture, and local grayscale features achieve the best overlap.

Practically, due to the presence of noise and variations in the mean brightness and contrast of the images, zero-mean normalized cross-correlation (NCC) is used to enhance the robustness of the displacement calculation. For a given displacement $(u,v)$, the mean grayscale values of the two images within the valid overlap region are
\begin{equation}
\bar{x}_{u,v} =
\frac{1}{|\Omega_{u,v}|}
\sum_{(i,j)\in\Omega_{u,v}} x_{ij},
\qquad
\bar{y}_{u,v} =
\frac{1}{|\Omega_{u,v}|}
\sum_{(i,j)\in\Omega_{u,v}} y_{i+u,j+v}.
\end{equation}

The corresponding normalized cross-correlation coefficient is
\begin{equation}
C_{\mathrm{NCC}}(u,v) =
\frac{
\sum_{(i,j)\in\Omega_{u,v}}
(x_{ij}-\bar{x}_{u,v})(y_{i+u,j+v}-\bar{y}_{u,v})
}{
\sqrt{
\sum_{(i,j)\in\Omega_{u,v}}(x_{ij}-\bar{x}_{u,v})^2
}
\sqrt{
\sum_{(i,j)\in\Omega_{u,v}}(y_{i+u,j+v}-\bar{y}_{u,v})^2
}
}.
\end{equation}

After normalization, $C_{\mathrm{NCC}}(u,v)$ ranges from $-1$ to $1$, and a value closer to $1$ indicates greater similarity between the two images at the corresponding displacement. By searching over a predefined two-dimensional displacement range $D$, the translational mismatch is determined from the location of the maximum NCC value:
\begin{equation}
(\hat{u},\hat{v}) =
\arg\max_{(u,v)\in D} C_{\mathrm{NCC}}(u,v).
\end{equation}

To assess the consistency between the two images after translational compensation, an MSE image is used to visualize the local grayscale difference. This MSE image is not obtained by reducing the difference between the two images to a single scalar. Instead, the squared error is calculated at each pixel and represented as a grayscale image. For two registered grayscale images $\mathbf{X}$ and $\mathbf{Y}$ of size $M \times N$, the squared error at row $i$ and column $j$ is
\begin{equation}
e_{ij} = (x_{ij}-y_{ij})^2.
\end{equation}

A larger value of $e_{ij}$ indicates a greater local grayscale difference between the two images. The scalar MSE, when required to quantify the overall difference between the two images, is calculated as
\begin{equation}
\mathrm{MSE}(\mathbf{X},\mathbf{Y}) =
\frac{1}{MN}\sum_{i=1}^{M}\sum_{j=1}^{N}e_{ij}.
\end{equation}

After obtaining the translational mismatch, the translational displacement is first compensated. The compensated image is then rotated over a set of trial angles, and the scalar MSE between the rotated image and the reference image is calculated at each angle. The angle that minimizes the MSE is identified as the rotational mismatch. For the chips with little displacement, the rotational mismatch is almost zero, which is negligible compared with the translational mismatch. For the chips with significant rotation, the translational displacement is always significantly $>$ 3.5 \textmu m (criteria for a successful transfer). In this case, a 2$\sim$3 \textmu m error in the manual measurement of the overall mismatch directly on the images will not significantly affect the evaluation of the transfer performance; therefore, the overall mismatch is measured directly on the images manually.

\clearpage

\begin{figure} 
	\centering
	\includegraphics[width=0.9\textwidth]{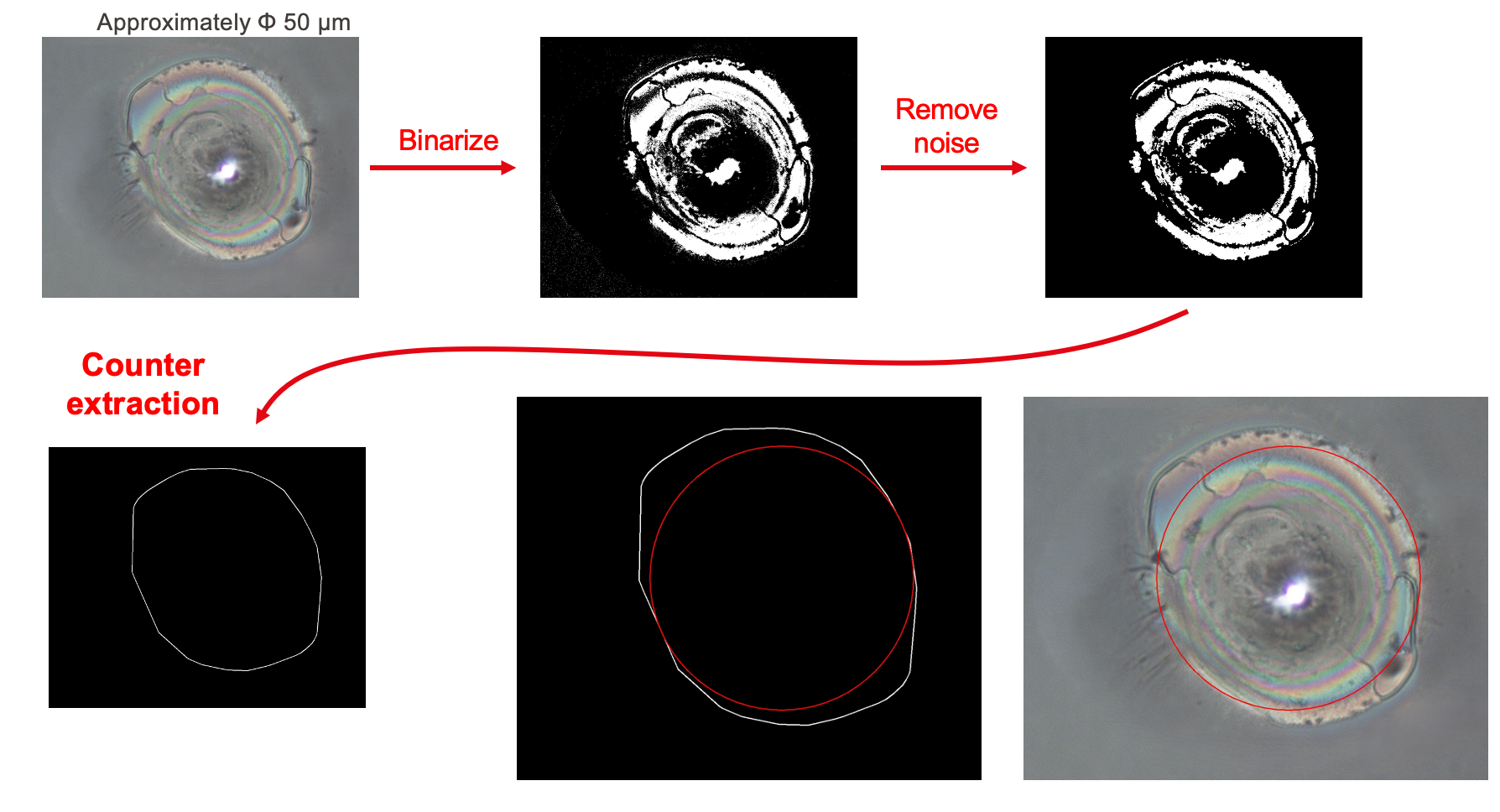} 

	\caption{\textbf{Counter extracting procedures for the ablation profile.}
		(A) Optical image of the ablation profile. (B) Binarized image of (A). (C) Applying noise removal on (B). (D) Extracting the counter outline based on the denoised picture. (E) Finding the largest inscribe circle of the extracted counter.}
	\label{fig:sup_counter} 
\end{figure}

\begin{figure} 
	\centering
	\includegraphics[width=0.4\textwidth]{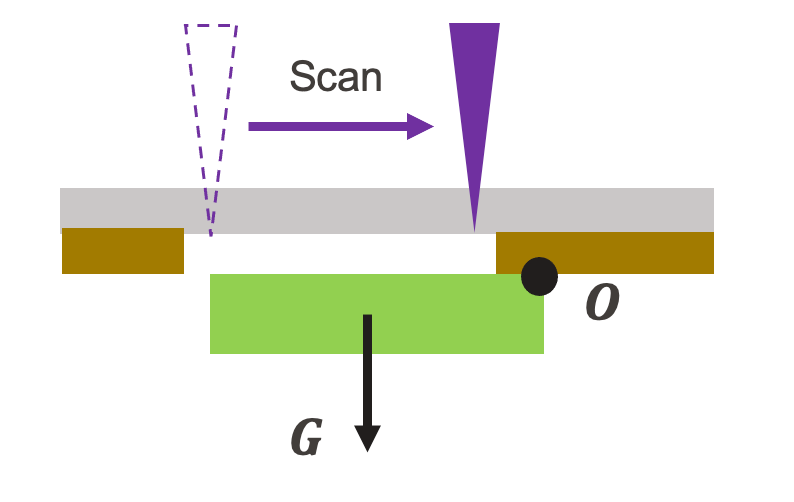} 

	\caption{\textbf{Illustration of the large-sized chips releasing mechanism.} When released from the side rather than the center of the chip, the chip will experience a torque that causes it to rotate and fall off the receiver, which is more likely to cause greater mismatch.}
	\label{fig:sup_gravity} 
\end{figure}

\begin{figure} 
	\centering
	\includegraphics[width=0.9\textwidth]{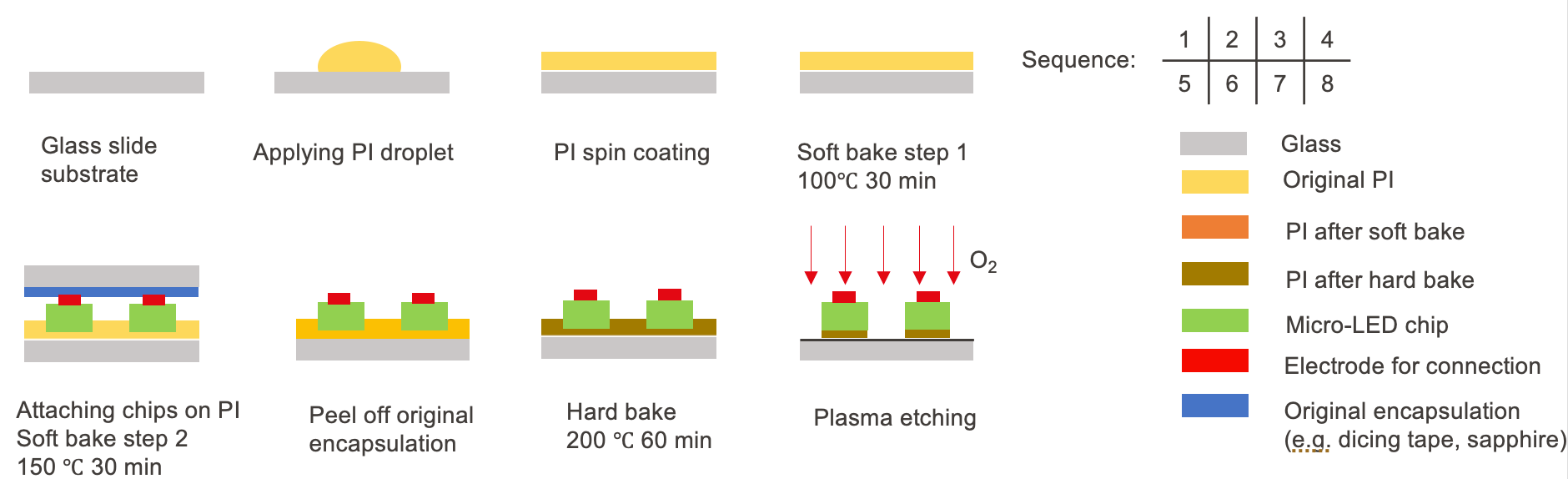} 

	\caption{\textbf{Schematic of the source substrate fabrication process.}}
	\label{fig:sup_source} 
\end{figure}

\begin{figure} 
	\centering
	\includegraphics[width=\textwidth]{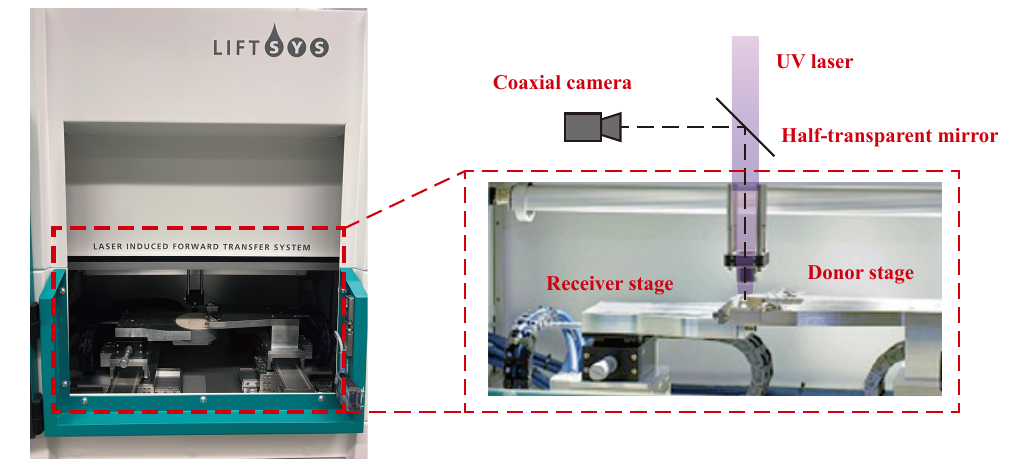} 

	\caption{\textbf{LIFT machine setup.}}
	\label{fig:sup_machine} 
\end{figure}

\end{document}